\documentclass[english,aps,prl,twocolumn,showpacs]{revtex4-1}
\usepackage[T1]{fontenc}
\usepackage[latin9]{inputenc}
\usepackage{amssymb}
\usepackage{graphicx}
\usepackage{esint}
\usepackage{babel}
\begin{document}

\title{Chiral anomaly and anomalous finite-size conductivity in graphene}

\author{Shun-Qing Shen}

\affiliation{Department of Physics, The University of Hong Kong, Pokfulam Road,
Hong Kong, China}

\author{Chang-An Li}

\affiliation{Department of Physics, The University of Hong Kong, Pokfulam Road,
Hong Kong, China}

\author{Qian Niu}

\affiliation{Department of Physics, University of Texas at Austin, Texas 78712,
USA}

\date{\today }
\begin{abstract}
Graphene is a monolayer of carbon atoms packed into a hexagon lattice
to host two pairs of massless two-dimensional Dirac fermions in the
absence of or with negligible spin-orbit coupling. It is known that
the existence of non-zero electric polarization in reduced momentum
space which is associated with a hidden chiral symmetry will lead
to the zero-energy flat band of zigzag nanoribbon. The Adler-Bell-Jackiw
chiral anomaly or non-conservation of chiral charges at different
valleys can be realized in a confined ribbon of finite width. In the
laterally diffusive regime, the finite-size correction to conductivity
is always positive and goes inversely with the square of the lateral
dimension $W$, which is different from the finite-size correction
inversely with $W$ from boundary modes. This anomalous finite-size
conductivity reveals the signature of the chiral anomaly in graphene,
and is measurable experimentally.
\end{abstract}
\maketitle
\emph{Introduction} - Graphene \cite{Neto09rmp} is a monolayer of
carbon atoms packed into a hexagon lattice to host two pairs of two-dimensional
Dirac fermions or four Dirac cones in the absence of or with negligible
spin-orbit coupling. Each pair of Dirac fermions have two Dirac cones
with opposite chirality, where the well-defined Berry phase $\pi$
or $-\pi$ rests around the Dirac cones when electron moves around
one Dirac point adiabatically \cite{Mikitik99prl,Suzuura02prl,Shen04prb,Young15prl}.
The chiral nature and topological properties of the Dirac fermions
in graphene have been extensively discussed over the past decade,
such as the anomalous integer quantum Hall effect \cite{ZhangYB05nat,Novoselov05nat},
Klein tunneling \cite{Katsnelson06np}, and valley physics \cite{Xiao12prl,Rycerz07np}.
Stimulated by the recent advances in three-dimensional Weyl semimetals
\cite{Volovik03book,Wan11prb,Xu11prl,Burkov11prl,Yang11prb,Xu15sci-TaAs,Lv15prx,ZhangCL16nc},
we come to re-examine the topological properties and chiral nature
of two-dimensional Dirac fermions in graphene.

Starting with the tight-binding model for electrons in graphene, it
was known that there exists the non-zero electric polarization in
reduced one-dimensional momentum space, which is associated with a
hidden chiral symmetry \cite{Ryu02prl}. This topological invariant
leads to the famous zero-energy flat band for a stripe of graphene
with zigzag boundary. For a zigzag nanoribbon of finite width, the
energy band dispersions become a series of discrete one-dimensional
bands, which are very similar to the Landau bands for the Weyl fermions
in a magnetic field, and the chiral currents exist along the boundary.
Here we show that the Adler-Bell-Jackiw chiral anomaly \cite{Adler69pr,Bell69Jackiw}
or non-conservation of chiral charge at two different valleys can
be realized by applying an electric field along the ribbon. The change
rate of chiral electrons at each valley is proportional to electric
field, but inversely to the lateral dimension $W$. In the laterally
diffusive regime where the mean free path is shorter than $W$, the
energy balance between the energy transfer between two valleys and
the Joule's heating of electric current gives a residual correction
to the finite-size conductivity inversely proportional to the square
of the lateral dimension $W$. This is opposite to the case that the
conductivity goes inversely with $W$ if the lateral modes can be
resolved or the mean free path is much bigger than $W$. This anomalous
behavior of finite-size effect indicates that a narrower ribbon may
have a larger finite-size conductance, manifesting a signature of
the chiral anomaly of two-dimensional Dirac fermions in graphene. 

\begin{figure}
\includegraphics[width=8.5cm]{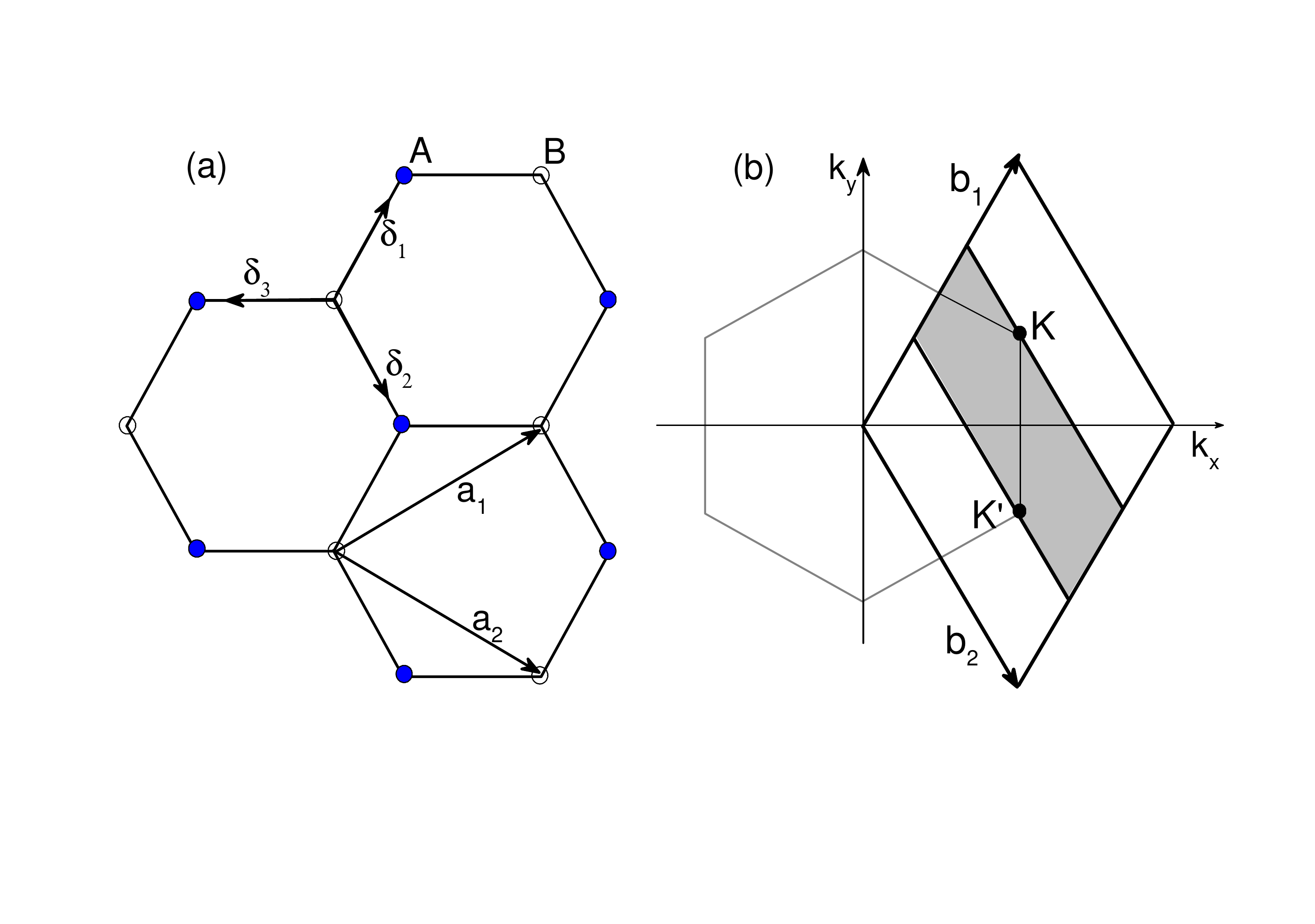}\caption{(a) The lattice unit vectors $\mathbf{a}_{1}$ and $\mathbf{a}_{2}$
in a unit cell of a hexagonal lattice of graphene. (b) The corresponding
Brillouin zone, and the Dirac cones are located at the $\mathbf{K}$
and $\mathbf{K}'$ points. The shadowed area indicates the region
with non-zero electric polarization.}
\end{figure}

\textit{Spinless graphene as a two-dimensional topological Dirac semimetal}
- Since the spin-orbit coupling in graphene is tiny \cite{Yao07prb},
we assume the spin degeneracy of electrons here. The energy dispersions
of electrons in graphene have been investigated extensively\cite{Neto09rmp,Wakabayashi10stam}.
To diagonalize the tight-biding model for graphene, a basis of two-component
``spinors'' of Bloch states constructed on the two sublattices $\mathbf{A}$
and $\mathbf{B}$ is introduced \cite{Neto09rmp}. Let $\delta_{1}$,
$\delta_{2}$ and $\delta_{3}$ be the displacements from a B site
to its three nearest-neighbor A sites as shown in Fig. 1(a). The lattice
vectors are chosen to be $\mathbf{a}_{1}=\sqrt{3}a(\frac{\sqrt{3}}{2},\frac{1}{2})$
and $\mathbf{a}_{2}=\sqrt{3}a(\frac{\sqrt{3}}{2},-\frac{1}{2})$.
The corresponding reciprocal lattice vectors are $\mathbf{b}_{1}=\frac{4\pi}{3a}(\frac{1}{2},\frac{\sqrt{3}}{2})$
and $\mathbf{b}_{2}=\frac{4\pi}{3a}(\frac{1}{2},-\frac{\sqrt{3}}{2})$.
In this representation the Hamiltonian becomes
\begin{equation}
H=-t\sum_{\mathbf{k}}\left(\begin{array}{cc}
0 & \phi(\mathbf{k})\\
\phi^{*}(\mathbf{k}) & 0
\end{array}\right),\label{eq:Hamiltonian}
\end{equation}
where $\sigma_{i}$ are the Pauli matrices and $\phi(\mathbf{k})=\sum_{n=1}^{3}e^{i\mathbf{k}\cdot(\delta_{n}-\delta_{1})}$.
The energy band dispersions are obtained as, 
\begin{equation}
E_{s}=-st\left|\phi(\mathbf{k})\right|,\label{eq:dispersion}
\end{equation}
where $s=\pm$. The two bands touch at the points of $\mathbf{K}=(\frac{2\pi}{3a},\frac{2\pi}{3\sqrt{3}a})=\frac{2}{3}\mathbf{b}_{1}+\frac{1}{3}\mathbf{b}_{2}$
and $\mathbf{K}'=(\frac{2\pi}{3a},-\frac{2\pi}{3\sqrt{3}a})=\frac{1}{3}\mathbf{b}_{1}+\frac{2}{3}\mathbf{b}_{2}$
in the Brillouin zone formed by the reciprocal lattice vectors $\mathbf{b}_{1}$
and $\mathbf{b}_{2}$ as shown in Fig. 1(b), and the dispersions become
linear in $k$ approximately measured from the two points. Thus two
Dirac cones are formed around $\mathbf{K}$ and $\mathbf{K}'$. The
density of states is linear in energy $E$ near the two Dirac points,
which is a key feature of a semimetal. The vector connecting $\mathbf{K}$
and $\mathbf{K}'$ is along $\delta\mathbf{K}=\mathbf{K}-\mathbf{K}'=\frac{1}{3}\left(\mathbf{b}_{1}-\mathbf{b}_{2}\right)$.
To explore the topological properties of the two Dirac cones, we take
$\mathbf{k}=k_{1}\mathbf{b}_{1}+k_{2}\mathbf{b}_{2}$ and consider
a specific direction by taking a specific value of $k_{1}$ (or $k_{2}$
) as a constant, and the model is reduced to one-dimensional along
the $\mathbf{b}_{2}$ (or $\mathbf{b}_{1}$) direction. When $k_{1}$
is away from the Dirac points, i.e., $k_{1}\neq1/3$ and $2/3$, the
reduced one-dimensional system always possesses a finite band gap.
This gap will close and re-open when $k_{1}$ is sweeping over one
Dirac point, which indicates that a topological phase transition may
occur in the process. 

Ryu and Hatsugai \cite{Ryu02prl} first realized that zero energy
modes in the zigzag ribbon of graphene are closely associated to a
hidden chiral symmetry for a reduced one-dimensional Hamiltonian in
a parameter space. Denote the eigenstates of $H$ by $\left|\phi_{s}\left(k_{x},k_{y}\right)\right\rangle =\frac{1}{\sqrt{2}}\left(1,s\phi(\mathbf{k})/\left|\phi(\mathbf{k})\right|\right)^{T}$with
the eigenvalue $E_{s}(\mathbf{k})$ in Eq. (\ref{eq:dispersion}).
In reduced one-dimensional momentum space, e.g., keeping $k_{1}$
constant, the electric polarization for the reduced one-dimensional
bands is defined as \cite{Xiao10rmp,Kingsmith93prb} 
\begin{equation}
P_{s}=-\frac{e\sqrt{3}a}{2\pi}\oint_{C}dk_{2}\left\langle \phi_{s}(k_{1},k_{2})\right|(i\partial_{k_{2}})\left|\phi_{s}(k_{1},k_{2})\right\rangle .\label{eq:winding-number}
\end{equation}
The electric polarization has its topological origin, and is quantized
to be $P_{s}=e\nu_{s}\sqrt{3}a$ where the integer $\nu_{s}=0$ or
1/2 with modulo 1 appears as a topological invariant for quantum transport.
It is found that $P_{s}=se\sqrt{3}a/2$ for $1/3<k_{1}<2/3$, and
$P_{s}=0$ otherwise. Thus the one-dimensional system is topologically
nontrivial when $1/3\leq k_{1}<2/3$. Therefore a spinless graphene
is a two-dimensional topological Dirac semimetal characterized by
a topological invariant, which is very similar to the three-dimensional
Weyl semimetal where a k-dependent Chern number is defined \cite{Hosur13Physique}.
According to the bulk-boundary correspondence, there exists a pair
of end modes near the two ends of an open and topological nontrivial
chain, and the energy of the end states should be zero due to the
chiral symmetry \cite{Ryu02prl}, just as the end modes in the Su-Schrieffer-Heeger
model \cite{Shen12book}. Correspondingly, the open boundary parallel
with the vector $\mathbf{b}_{1}$ (or $\mathbf{b}_{2}$) is the zigzag
boundary of the lattice. Each $k_{1}$ in the shadow regime will have
a pair of zero energy modes near two edges according to the the bulk-edge
correspondence and the particle-hole symmetry in Eq. (1), and a flat
band is formed for a zigzag boundary of the graphene lattice. This
demonstrates that the famous flat band of graphene with a zigzag boundary
condition has its topological origin related to the non-zero electric
polarization or topological invariant, just as the Fermi arc in three-dimensional
Weyl semimetals \cite{Lu15Weyl-shortrange}. Unlike that a magnetic
monopole is located at a three-dimensional Weyl node, a well-defined
$\pi$ Berry phase exists around the Dirac point in two dimensions
\cite{Suzuura02prl,Mikitik99prl,Shen04prb}.

\textit{Chiral anomaly} - It is well known that the chiral charge
conservation may be violated for Weyl fermions in three dimensions.
Nielsen and Ninomiya \cite{Nielsen83plb} proposed to simulate the
chiral anomaly in one-dimensional chiral bands by using the lowest
Landau band of a three-dimensional semimetal in the presence of a
magnetic field. Applying an external electric field along the magnetic
field will drive electrons flowing from one node to another one, which
leads to the so-called Adler-Bell-Jackiw chiral anomaly \cite{Adler69pr,Bell69Jackiw}.
Consequently, it is expected that the longitudinal magnetoresistivity
becomes extremely strong, and negative \cite{Nielsen83plb}. If the
electric field is normal to the magnetic field, the anomaly will disappear.
Recently, thanks to the discovery of a number of realistic materials
of topological semimetal \cite{Xu15sci-TaAs,Lv15prx}, there is growing
passion on their electronic transport and signatures of the chiral
anomaly \cite{Xiong15sci,ZhangCL16nc,LiH16nc,LiQ-16np}. Inspired
by the advances in three-dimensional Weyl semimetals, it becomes an
appealing issue whether or not there also exists the chiral anomaly
or its signature in two-dimensional Dirac semimetal or graphene \cite{Young15prl}.
A perpendicular magnetic field can induce discrete Landau bands, but
it is always perpendicular to the electric field or electric current
in graphene, and an in-plane magnetic field cannot produce Landau
bands. Thus it is obvious that there is no chiral anomaly in graphene
in the presence of both magnetic and electric fields, opposite to
the three-dimensional case. However early works on quasi-one-dimensional
graphene nanoribbon has shown that they may support the formation
of one-dimensional chiral bands due to the finite-size effect of quantum
confinement \cite{Peres06prb}.

\begin{figure}
\includegraphics[width=7.5cm]{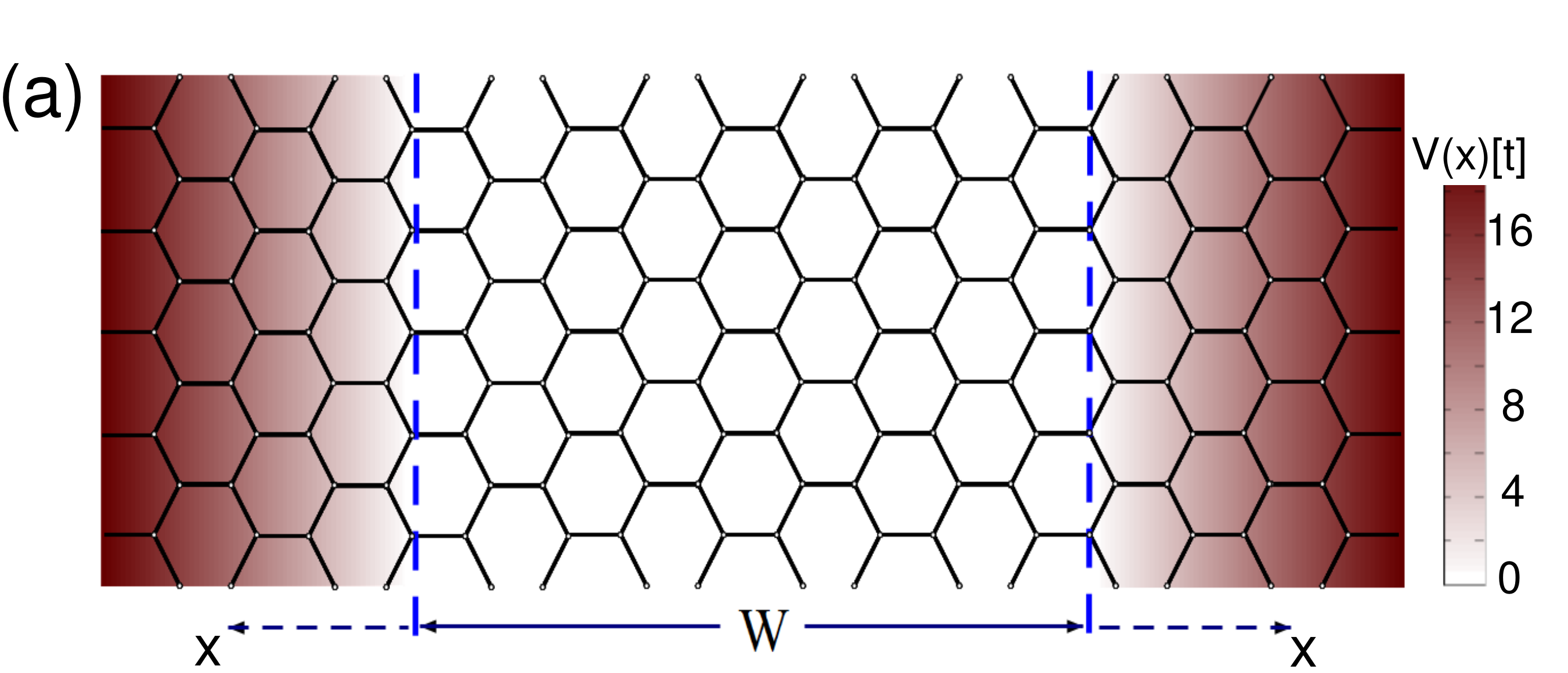}

\includegraphics[width=7.5cm]{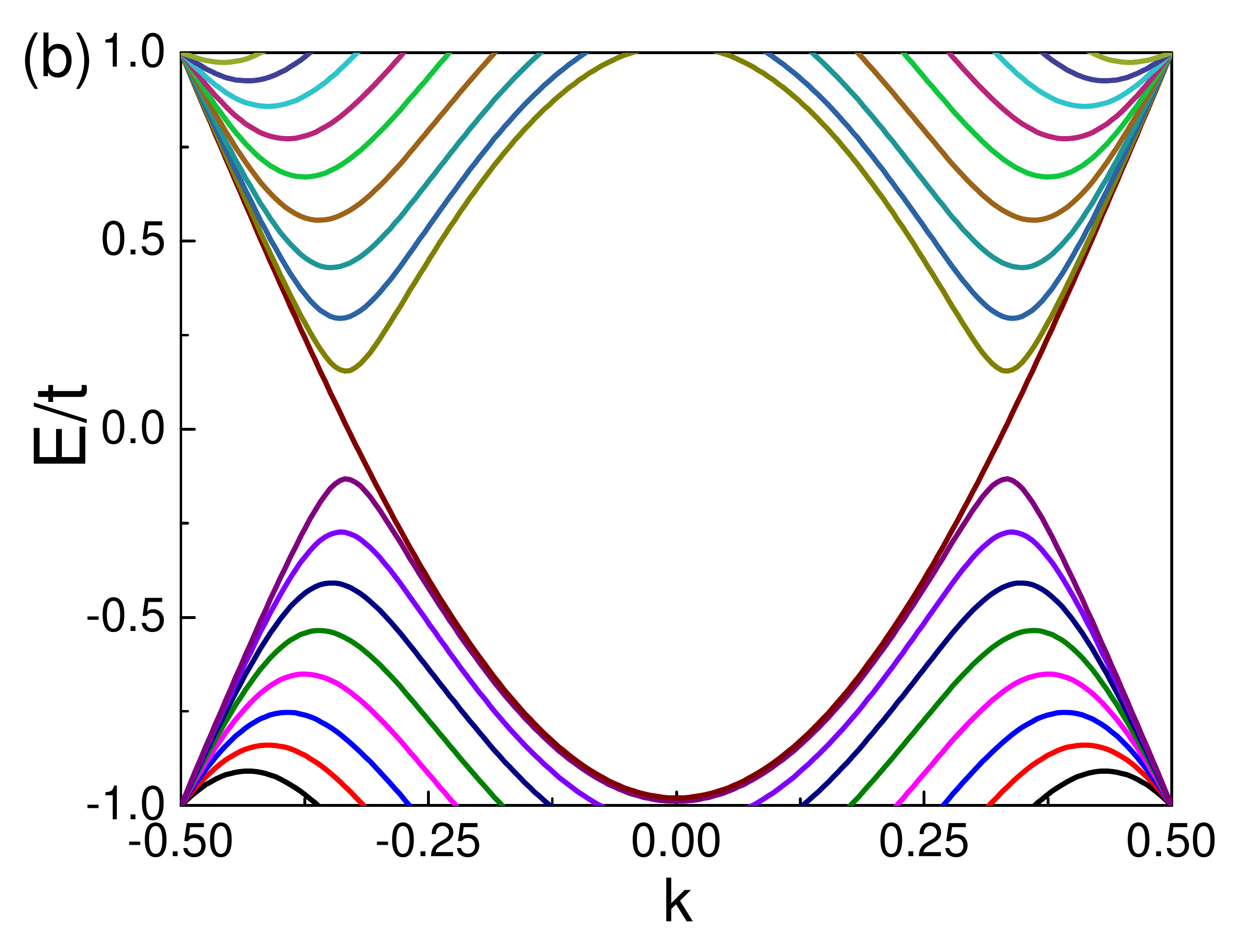}

\caption{(a) Schematic of a carrier-guiding graphene sheet encapsulated by
two confining gates or potentials along the zigzag direction. The
shadow area means a confining potential $V(x)$ is added, say $V(x)=V_{0}\left|x\right|$
where $x$ is the position measured from the potential boundary, i.e.,
the dashed lines. The potential is zero in the white regime. (b) The
energy band dispersions near two Dirac points. $k$ is the wave vector.
The parameters are $N=20$ , $V_{0}=\frac{3t}{a}$, and $a_{0}=\sqrt{3}a$.}
 \label{dispersion-chiral}
\end{figure}

Consider a monolayer graphene sheet encapsulated by two confining
gates along the zigzag direction to realize the quasi-one-dimensional
ballistic channels as shown in Fig. 2. The width of the free channel
is $W=(\frac{3}{2}N-1)a$, where $N\geq2$ is the site number of one
sublattice along the unit vector. Without loss of generality, a confining
potential is taken to be $V(x)=V_{0}\left|x\right|$ where $x$ is
the position measured away from the dashed line or the potential boundary
or as shown in Fig. 2(a). $V_{0}$ represents the strength of the
confining potential. It is worth of emphasizing that the finial result
in this paper is not so sensitive to the form of the confining potential.
Due to the effect of quantum confinement, the energy dispersion evolves
into a set of one-dimensional chiral bands. Modes $n=1,2,3,\cdots,N$
lie in the right valley whereas modes $n=-1,-2,\cdots,-N$ lie in
the left valley. At each valley, the energy spacing for different
non-zero energy bands is approximately $\triangle=\frac{3\pi}{2}\frac{ta}{W}$,
and the gap between conduction band and valence band is about $3\Delta$.
Within the band gap, the zeroth mode, $n=0$, with nearly linear dispersion
crossing two the Dirac points at $k=\pm1/3$ with opposite velocities,
$v=\pm\frac{3}{2}ta/\hbar$. Take the periodic boundary condition
along the vertical direction. The energy band dispersion is shown
in Fig. \ref{dispersion-chiral}b. The energy dispersions are split
into a series of energy bands due to the confining effect of the potential,
which are very similar to the Landau bands induced by an external
magnetic field in graphene \cite{Rammal85jp}, and also those in topological
Weyl semimetals \cite{Lu15Weyl-shortrange}. It is noted that the
shape and position of $n=0$ mode depends on the strength of potential.
When $V_{0}\rightarrow\infty$, it is similar to a zigzag nanoribbon,
in which the modes will be flattened \cite{Peres06prb,Rycerz07np}. 

In the absence of an external electric field, if the Fermi level is
near $E_{F}=0$, the effective velocities at the points of the Fermi
level are opposite to each other, the right one is positive and the
left one is negative, indicating the formation of chiral states. Applying
an external electric field will drive electrons flowing from one valley
to the other. The change rate of the density of charge carriers at
one valley $s$($=\pm$) is $\frac{dn_{s}}{dt}=\frac{s}{2\pi W}\frac{2\pi}{a_{0}}\frac{dk}{dt}$,
and $\frac{d}{dt}\left(k\frac{2\pi}{a_{0}}\right)=\frac{eE}{\hbar}$.
Thus the change rate at each valley is 
\begin{equation}
\frac{dn_{s}}{dt}=s\frac{e}{h}\frac{E}{W},\label{eq:change-rate}
\end{equation}
indicating the violation of the charge conservation at each valley,
although the total charges of the whole system is conserved, $d(n_{+}+n_{-})/dt=0$.
Thus the chiral anomaly of two-dimensional Dirac fermions can also
be realized, as an analog of three-dimensional Weyl fermions in parallel
electric and magnetic fields \cite{Nielsen83plb}. For three-dimensional
Weyl fermions, the changing rate of density at each valley is proportional
to $\mathbf{B}\cdot\mathbf{E}$ ($\mathbf{B}$ is the magnetic field
and \textbf{E} is the electric field), and the chiral anomaly disappears
when the electric field is normal to the magnetic field. Here the
anisotropy of chiral anomaly in graphene is determined by the orientation
of the confining potential. If the confining potential is along the
armchair direction $\mathbf{a}_{0}$, i.e., along the direction of
$\mathbf{a}_{1}+\mathbf{a}_{2}$, the chiral band crossing the conduction
and valence bands may disappear, and consequently, so does the chiral
anomaly. For a specific orientation of confining potential between
the armchair and zigzag orientation, $\mathbf{a}=\cos\theta\mathbf{a}_{2}+\sin\theta\mathbf{a}_{0}$,
where $\theta$ varies from 0 to $\pi/2$, the change rate will be
modified by a factor $\cos\theta$ , which approaches zero when $\theta=\pi/2$. 

It is noted that the ferromagnetism may appear along the zigzag edge
of a graphene nanoribbon \cite{Son-Nature-06,Tao-11np}. Due to the
fact that the dispersions still connect two valleys, the current mechanism
of the chiral anomaly will survive when the Fermi level crosses the
the edge dispersions. 

\textit{Signature of chiral anomaly} - Now we come to explore the
measurable signature of chiral anomaly in graphene. It is known that
the chiral anomaly of Weyl fermions generates the chiral magnetic
effect \cite{Vilenkin80prd,Nielsen83plb,Fukushima08prd}. Its magnetoconductivity
is positive and quadratic in magnetic field B \cite{LiQ-16np}. As
the change rate of Dirac fermions at each valley in graphene is proportional
to $1/W$, a finite-size conductivity is expected to reveal a signature
of chiral anomaly for Dirac fermions in graphene. Consider the relation
between the density of charge carriers $n_{\pm}$ and the chemical
potential $\mu_{\pm}$, $n_{\pm}=\frac{1}{4\pi}\frac{\mu_{\pm}^{2}}{\left(v_{F}\hbar\right)^{2}}$
at each valley. Without loss of generality, assume the chemical potentials
are positive, i.e., in the electron doped case. In the presence of
an electric field, electrons will flow from the right valley to the
left valley, and cause a difference of chemical potentials, $\delta\mu=\mu_{+}-\mu_{-}.$
However, the process will be balanced by the inter-valley scattering
between the left and right moving electrons. Consider the lateral
diffusive regime where the mean free path is shorter than the width,
and denote the inter-valley scattering time by $\tau_{v}$. The equation
for the chiral anomaly is reduced to 
\begin{equation}
\frac{d(n_{+}-n_{-})}{dt}=2\frac{eE}{hW}-\frac{n_{+}-n_{-}}{\tau_{v}}.
\end{equation}
In a steady state, the solution to the equation is $n_{+}-n_{-}=2\frac{eE}{hW}\tau_{v}$.
Thus the difference of the chemical potentials at two valley is $\delta\mu=\frac{2\pi\left(v_{F}\hbar\right)^{2}}{\mu}\frac{2eE}{hW}\tau_{v}$
(assume $\mu\neq0$)\cite{note-1}. As the Fermi levels at two valleys
are different, and the change rate of electrons is finite due to the
chiral anomaly, the energy cost for transferring the electrons from
one valley to the other is simply the product of the change rate of
electrons at each valley and the energy difference between the two
valleys. This energy must be extracted from the Joule's heating in
the presence of a charge current, and the energy balance gives $jE=2\frac{eE}{hW}\delta\mu$
\cite{Nielsen83plb}, where the factor 2 comes from the spin degeneracy
in graphene. Thus the solution to the current is
\begin{equation}
j=\frac{2e}{hW}\delta\mu=2\frac{e^{2}v_{F}^{2}\tau_{v}}{\pi\mu W^{2}}E
\end{equation}
Consequently the conductivity correction due to the chiral anomaly
is $\sigma_{W}=\frac{2e^{2}v_{F}^{2}\tau_{v}}{\pi\mu W^{2}}$, which
is proportional to $1/W^{2}$. It is an anomalous behavior that the
conductivity correction increases with decreasing the width in $W^{2}$.
The inter-valley scattering time $\tau_{v}$ depends on the scattering
mechanism. If the scattering potential is short ranged of the form
$V=v_{0}\delta(\mathbf{r})$, the inter-valley scattering time is
given by $\tau_{v}\sim\frac{4v_{F}}{n_{i}v_{0}^{2}}\frac{1}{\left|\delta\mathbf{K}\right|}$
where $n_{i}$ is the impurity density \cite{Shon-98jpsj}. Thus $\sigma_{W}$
is proportional to the square root of the density of charge carriers,
$\sigma_{W}\propto1/\sqrt{n}$. To estimate the order of the finite
size correction, we take $\mu=\tilde{\mu}$eV, $W=\tilde{W}$nm, $l=v_{F}\tau_{v}=\tilde{l}$nm,
and $v_{F}=10^{6}\mathrm{m/s}$, and have $\sigma_{W}\simeq2.63\frac{\tilde{l}}{\tilde{\mu}\tilde{W}^{2}}\frac{e^{2}}{h}$
. $\tilde{\mu}\sim0.1$ for the charge density $n\sim10^{12}/\mathrm{cm}^{2}$.
The intervalley scattering length is much longer than the mean free
path in graphene \cite{Shon-98jpsj,Morpurgo-06prl}. Thus for a sample
with scattering length of several tens of nanometer, the correction
is measurable for $W$ up to several tens or even hundred of nanometers.
The conventional Drude conductivity of graphene $\sigma_{0}$ was
found to proportional to the density of charge carriers, $n$, while
reaches at a minimal conductivity at the charge neutral point \cite{Shon-98jpsj,Nomura-07prl,Cheianov-06prl,Adam-07pnas}.
The conductivity is always of the order of $e^{2}/h$ and has a weak
dependence on size \cite{Lewenkopf-08prb,Neto09rmp}. The quantized
conductance in steps of $4e^{2}/h$ has been measured in a graphene
nanoribbon \cite{Kim-16np}. For wires or ribbons without boundary
modes, the conductivity usually has a negative correction due to excess
scattering from the boundary which is inversely proportional to $W$.
In the presence of a boundary mode, the correction can become positive,
but still goes inversely with $W$. However, the anomalous correction
here is always positive and goes inversely with the square of $W$. 

\textit{Summary }- In short, we propose that the chiral anomaly of
two-dimensional Dirac fermions can be realized in a quasi-one-dimensional
monolayer graphene, and its signature can be detected by measuring
the finite-size conductivity. It is necessary to emphasize that we
consider the laterally diffusive regime where the mean free path is
shorter than the width. When the mean free path is much larger than
the width $W$, the story will be very different. It is believed that
this property is associated with the chiral symmetry of the Dirac
fermions in the reduced one-dimensional momentum space, and the formation
of chiral currents of the system. Chiral anomaly is a purely quantum
mechanical effect for Weyl fermions. It is widely accepted that the
positive magnetoconductivity or negative magnetoresistance in topological
Weyl semimetals is a signature of chiral anomaly. The anomalous finite-size
conductivity in graphene may provide an alternative way to detect
this quantum mechanical effect in solid. 

This work was supported by the Research Grant Council, University
Grants Committee, Hong Kong under Grant No. 17304414 and HKU3/CRF/13G.
SQ would like to thank Di Xiao, Jian Li, Hai-Zhou Lu and Jian-Hui
Zhou for helpful discussions. This research is conducted in part using
the HKU ITS research computing facilities that are supported in part
by the Hong Kong UGC Special Equipment Grant (SEG HKU09).


\begin{thebibliography}{10}
\bibitem{Neto09rmp}A. H. Castro Neto, F. Guinea, N. M. R. Peres,
K. S. Novoselov, and A. K. Geim, Rev. Mod. Phys. 81, 109 (2009).

\bibitem{Mikitik99prl}G. P. Mikitik and Y. V. Sharlai, Phys. Rev.
Lett. 82, 2147 (1999).

\bibitem{Suzuura02prl}H. Suzuura and T. Ando, Phys. Rev. Lett. 89,
266603 (2002).

\bibitem{Shen04prb} S. Q. Shen, Phys. Rev. B 70, 081311(R) (2004).

\bibitem{Young15prl}S. M. Young and C. L. Kane, Phys. Rev. Lett.
115, 126803 (2015).

\bibitem{ZhangYB05nat} Y. B. Zhang, Y. W. Tan, H. L. Stormer, and
P. Kim, Nature 438, 201 (2005).

\bibitem{Novoselov05nat}K. S. Novoselov, A. K. Geim, S. V. Morozov,
D. Jiang, M. I. Katsnelson, I. V. Grigorieva, S. V. Dubonos, and A.
A. Firsov, Nature 438, 197 (2005).

\bibitem{Katsnelson06np}M. I. Katsnelson, K. S. Novoselov, and A.
K. Geim, Nat. Phys. 2, 620 (2006).

\bibitem{Xiao12prl}D. Xiao, G.-B. Liu, W. Feng, X. Xu, and W. Yao,
Phys. Rev. Lett. 108, 196802 (2012).

\bibitem{Rycerz07np}A. Rycerz, J. Tworzydlo, and C. W. J. Beenakker,
Nat. Phys. 3, 172 (2007).

\bibitem{Volovik03book}G. E. Volovik, The Universe in a Helium Droplet
(Clarendon Press, Oxford, 2003).

\bibitem{Wan11prb}X. Wan, A. M. Turner, A. Vishwanath, and S. Y.
Savrasov, Phys. Rev. B 83, 205101 (2011).

\bibitem{Xu11prl}G. Xu, H. M. Weng, Z. J. Wang, X. Dai, and Z. Fang,
Phys. Rev. Lett. 107, 186806 (2011).

\bibitem{Burkov11prl}A. A. Burkov and L. Balents, Phys. Rev. Lett.
107, 127205 (2011).

\bibitem{Yang11prb}K. Y. Yang, Y. M. Lu, and Y. Ran, Phys. Rev. B
84, 075129 (2011).

\bibitem{Xu15sci-TaAs}S. Y. Xu, I. Belopolski, N. Alidoust, M. Neupane,
G. Bian, C. L. Zhang, R. Sankar, G. Q. Chang, Z. J. Yuan, C. C. Lee,
et al., Science 349, 613 (2015).

\bibitem{Lv15prx}B. Q. Lv, H. M. Weng, B. B. Fu, X. P. Wang, H. Miao,
J. Ma, P. Richard, X. C. Huang, L. X. Zhao, G. F. Chen, et al., Phys.
Rev. X 5, 031013 (2015).

\bibitem{ZhangCL16nc}C. L. Zhang, S.-Y. Xu, I. Belopolski, Z. Yuan,
Z. Lin, B. Tong, G. Bian, N. Alidoust, C. C. Lee, S. M. Huang, et
al., Nat Commun. 7, 10735 (2016).

\bibitem{Ryu02prl}S. Ryu and Y. Hatsugai, Phys. Rev. Lett. 89, 077002
(2002).

\bibitem{Adler69pr}S. L. Adler, Phys. Rev. 177, 2426 (1969).

\bibitem{Bell69Jackiw}J. S. Bell and R. Jackiw, Il Nuovo Cimento
A 60, 47 (1969).

\bibitem{Yao07prb}Y. Yao, F. Ye, X. L. Qi, S. C. Zhang, and Z. Fang,
Phys. Rev. B 75, 041401 (2007). 

\bibitem{Wakabayashi10stam}K. Wakabayashi, K. ichi Sasaki, T. Nakanishi,
and T. Enoki, Science and Technology of Advanced Materials 11, 054504
(2010).

\bibitem{Xiao10rmp}D. Xiao, M. C. Chang, and Q. Niu, Rev. Mod. Phys.
82, 1959 (2010).

\bibitem{Kingsmith93prb}R. D. King-Smith and D. Vanderbilt, Phys.
Rev. B 47, 1651 (1993).

\bibitem{Hosur13Physique}P. Hosur and X. Qi, C. R. Physique 14, 857
(2013).

\bibitem{Shen12book}S. Q. Shen, Topological Insulators (Springer-Verlag,
Berlin Heidelberg, 2012). 

\bibitem{Lu15Weyl-shortrange}H. Z. Lu, S. B. Zhang, and S. Q. Shen,
Phys. Rev. B 92, 045203 (2015).

\bibitem{Nielsen83plb}H. B. Nielsen and M. Ninomiya, Physics Letters
B 130, 389 (1983).

\bibitem{Xiong15sci}J. Xiong, S. K. Kushwaha, T. Liang, J. W. Krizan,
M. Hirschberger, W. Wang, R. J. Cava, and N. P. Ong, Science 10, 1126
(2015).

\bibitem{LiH16nc}H. Li, H. He, H.-Z. Lu, H. Zhang, H. Liu, R. Ma,
Z. Fan, S. Q. Shen, and J. Wang, Nat Commun. 7, 10301 (2016).

\bibitem{LiQ-16np}Q. Li, D. E. Kharzeev, C. Zhang, Y. Huang, I. Pletikosic,
A. V. Fedorov, R. D. Zhong, J. A. Schneeloch, G. D. Gu, and T. Valla,
Nat. Phys. 12, 550 (2016).

\bibitem{Peres06prb}N. M. R. Peres, A. H. Castro Neto, and F. Guinea,
Phys. Rev. B 73, 195411 (2006).

\bibitem{Son-Nature-06}Y. W. Son, M. L. Cohen, and S. G. Louie, Nature
444, 347 (2006).

\bibitem{Tao-11np}C. Tao, et al., Nat. Phys. 7, 616 (2011)). 

\bibitem{Rammal85jp}R. Rammal, J. Phys. France 46, 1345 (1985).

\bibitem{Vilenkin80prd}A. Vilenkin, Phys. Rev. D 22, 3080 (1980).

\bibitem{note-1}For $\mu=0$, $\delta\mu=2v_{F}\hbar\sqrt{4\pi\frac{eE}{hW}\tau_{v}}$,
and the charge current is $j=\frac{2ev_{F}}{\pi W}\sqrt{4\pi\frac{eE}{hW}\tau_{v}}$.
However at this point other effects such as disorders or fluctuations
become dominant to produce a minimal conductivity. 

\bibitem{Fukushima08prd}K. Fukushima, D. Kharzeev and H. Warringa,
Phys. Rev. D 78, 074033 (2008).

\bibitem{Shon-98jpsj}N. H. Shon and T. Ando, J. Phys. Soc. Jpn. 67,
2421 (1998).

\bibitem{Morpurgo-06prl}A. F. Morpurgo and F. Guinea, Phys. Rev.
Lett. 97, 196804 (2006).

\bibitem{Nomura-07prl}K. Nomura and A. H. MacDonald, Phys. Rev. Lett.
98, 076602 (2007).

\bibitem{Cheianov-06prl}V. V. Cheianov and V. I. Falko, Phys. Rev.
Lett. 97, 226801 (2006).

\bibitem{Adam-07pnas}S. Adam, E. H. Hwang, V. M. Galitski and S.
Das Sarma, Proc. Natl. Acad. Sci. USA, 104, 18392 (2007).

\bibitem{Lewenkopf-08prb}C. H. Lewenkopf, E. R. Mucciolo, and A.
H. Castro Neto, Phys. Rev. B 77, 081410 (2008).

\bibitem{Kim-16np}M. Kim, J. H. Choi, S. H. Lee, K. Watanabe, T.
Taniguchi, S. H. Jhi and H. J. Lee, Nat. Phys. 12, 1022 (2016).
\end{thebibliography}
\end{document}